\documentclass[aps,prx,amsmath,amssymb,reprint,nofootinbib]{revtex4-1}\usepackage{graphicx}
\usepackage[uline]{hhtensor}
\usepackage[caption=false]{subfig}
\usepackage{mathtools}
\usepackage{cancel}
\usepackage{textcomp}
\newcommand\HH{\mathbf{H}^\omega}
\newcommand\EE{\mathbf{E}^\omega}
\newcommand\NN{\mathbf{N}^\omega}
\newcommand\MM{\mathbf{M}^\omega}
\newcommand\hatNN{\mathbf{\hat{N}}^\omega}
\newcommand\hatMM{\mathbf{\hat{M}}^\omega}

\newcommand\TT{\mathbf{T}}
\newcommand\Ajmp{\mathbf{A}_{jm+}^\omega}
\newcommand\Ajmm{\mathbf{A}_{jm-}^\omega}
\newcommand\hatAjmp{\mathbf{\hat{A}}_{jm+}^\omega}
\newcommand\hatAjmm{\mathbf{\hat{A}}_{jm-}^\omega}

\newcommand\aj{a_j^\omega}
\newcommand\bj{b_j^\omega}
\newcommand\alphaj{\alpha_{j\pm}^\omega}
\newcommand\betaj{\beta_{j\pm}^\omega}
\newcommand\alphajp{\alpha_{j+}^\omega}
\newcommand\betajp{\beta_{j+}^\omega}

\newcommand\ed{\mathbf{p}^\omega}
\newcommand\md{\mathbf{m}^\omega}

\newcommand\dE{{\matr{\alpha}}_{pE}^\omega}
\newcommand\dH{{\matr{\alpha}}_{pH}^\omega}
\newcommand\mE{{\matr{\alpha}}_{mE}^\omega}
\newcommand\mH{{\matr{\alpha}}_{mH}^\omega}

\newcommand\rrp{\mathbf{r_0}}
\newcommand\rr{\mathbf{r}}
\newcommand\T{\matr{T}^\omega}
\newcommand\SSS{\matr{S}^\omega}
\newcommand\Cinv{\matr{C}^{-1}}
\newcommand\C{\matr{C}}

\newcommand\cd{\begin{bmatrix}c^\omega_{1-1}\\c^\omega_{10}\\c^\omega_{11}\\d^\omega_{1-1}\\d^\omega_{10}\\d^\omega_{11}\end{bmatrix}}
\newcommand\ab{\begin{bmatrix}a^\omega_{1-1}\\a^\omega_{10}\\a^\omega_{11}\\b^\omega_{1-1}\\b^\omega_{10}\\b^\omega_{11}\end{bmatrix}}

\newcommand\Tnn{\T_{NN}}
\newcommand\Tnm{\T_{NM}}
\newcommand\Tmn{\T_{MN}}
\newcommand\Tmm{\T_{MM}}

\newcommand\Tpp{\T_{++}}
\newcommand\Tmimi{\T_{--}}
\newcommand\Tmip{\T_{-+}}
\newcommand\Tpmi{\T_{+-}}

\newcommand\vhat{\mathbf{\hat{v}}}
\newcommand\rhat{\mathbf{\hat{r}}}
\newcommand{\mupv}{\vec{\mu}^\omega_+\left(\vhat\right)}
\newcommand{\mumv}{\vec{\mu}^\omega_-\left(\vhat\right)}
\newcommand{\mupvz}{\vec{\mu}^\omega_+\left(\zhat\right)}
\newcommand{\mumvz}{\vec{\mu}^\omega_-\left(\zhat\right)}

\newcommand{\alphav}{\vec{\alpha}^\omega}
\newcommand\zhat{\mathbf{\hat{z}}}
\newcommand\xhat{\mathbf{\hat{x}}}
\newcommand\yhat{\mathbf{\hat{y}}}
\newcommand\ehatp{\mathbf{\hat{e}_1}}
\newcommand\ehatz{\mathbf{\hat{e}_0}}
\newcommand\ehatm{\mathbf{\hat{e}_{-1}}}
\newcommand\ehatmm{\mathbf{\hat{e}_{m}}}
\newcommand\wvec{\mathbf{w}}

\newcommand\Tr[1]{\textrm{Tr}\left\{#1\right\}}
\newcommand\RAp{\textrm{Ro}\left\{\matr{A}_+^\omega\right\}}
\newcommand\RAm{\textrm{Ro}\left\{\matr{A}_-^\omega\right\}}

\newcommand\Eye[1]{\matr{I}_{#1\times#1}}
\newcommand\Zero[2]{\matr{0}_{#1\times#2}}

\newcommand\alphawithzeros{\begin{bmatrix}\alphav_+(\vhat)\\0\end{bmatrix}}
\begin{document}
\title{Computation of electromagnetic properties of molecular ensembles}
\author{Ivan Fernandez-Corbaton}\email{ivan.fernandez-corbaton@kit.edu}\affiliation{Institute of Nanotechnology, Karlsruhe Institute of Technology, 76021 Karlsruhe, Germany}
\author{Carsten Rockstuhl}\affiliation{Institut f\"ur Theoretische Festk\"orperphysik, Karlsruhe Institute of Technology, 76131 Karlsruhe, Germany}
\affiliation{Institute of Nanotechnology, Karlsruhe Institute of Technology, 76021 Karlsruhe, Germany}
\author{Wim Klopper}\affiliation{Theoretical Chemistry Group, Institut for Physical Chemistry, Karlsruhe Institute of Technology, 76131 Karlsruhe, Germany}
\affiliation{Institute of Nanotechnology, Karlsruhe Institute of Technology, 76021 Karlsruhe, Germany}
\begin{abstract}
We establish a link between quantum mechanical molecular simulations and the transfer matrix of a molecule. The transfer matrix (T-matrix) of an object provides a complete description of its electromagnetic response. Once the T-matrices of the individual components of an ensemble are known, the electromagnetic response of the ensemble can be efficiently computed. This holds for arbitrary arrangements of large number of molecules, as well as for periodic arrays. We provide T-matrix based formulas for computing traditional chiro-optical properties like Circular Dichroism and Oriented Circular Dichroism, and also for quantifying electromagnetic duality and electromagnetic chirality, two properties that are fundamentally related to chiral interactions, and also technologically relevant. The formulas are valid for light-matter interactions of arbitrary high multipolar orders. We exemplify our approach by first computing the T-matrix of a cross-like arrangement of four copies of a chiral molecule from the time-dependent Hartree-Fock theory simulation data of the individual molecule, and then computing the aforementioned electromagnetic properties of both the cross and the individual molecule. The link that we establish is a necessary step towards obtaining T-matrix based constitutive relations of general bulk molecular materials from quantum mechanical simulations of their molecular constituents. 
\end{abstract}
\maketitle
Nanoscience is an interdisciplinary endeavor. The study of matter, radiation, and their interaction at tiny spatial scales involves many different fields in physics and chemistry. Moreover, engineering is eventually needed to convert the scientific advances into technological ones. Each discipline has its own methods, which difficults cross-talk and hinders progress. The molecular-based design of discrete objects and bulk materials for specific electromagnetic functions is a prominent example of interdisciplinary work where chemistry, physics, and engineering need to come together. It is also an example of one of the main challenges in nanotechnology: Scale heterogeneity \cite{Lieber2003}. It is hence desirable to develop approaches that connect methodologies across the different disciplines and length scales in nano and mesoscopic science and technology. One of the objectives of such approaches, which can open up opportunities in computational material science, is to predict the observable electromagnetic properties of macroscopic functional devices from descriptions at the molecular level. 

At the smallest length scales, quantum mechanical simulation methods based on, for example, time-dependent density functional theory (TD-DFT), time-dependent Hartree-Fock theory (TD-HF), or linear-response coupled-cluster (LR-CC) theory are used in computational chemistry to obtain the electromagnetic response of a single molecule~\cite{Bast2011,Helgaker2012,nrsbook}. One limitation of this kind of simulations is their computational complexity which renders them impractical when the system size goes beyond 10$^4$ atoms~\cite{Zuehlsdorff2015,Seibert2017}. In physics and engineering, the transfer matrix (T-matrix) approach  \cite{Waterman1965} is a very common tool for computing the electromagnetic responses of single and composite objects. Given the T-matrices of the individual constituents, the joint response of a composite system like random media, and periodic and aperiodic arrays, can be computed efficiently (see the references in \cite[Secs. 2.6 and 2.7]{Mishchenko2016}). In this approach, the computation of the individual T-matrices requires a basic model for light-matter interactions. The model is typically the macroscopic Maxwell equations featuring the constitutive relations: Electric permittivity, magnetic permeability, etc. The main limitation of this approach is that at small enough scales the effective field description implicit in the constitutive relations ceases to be a good approximation \cite[Chap. 6.6]{Jackson1998}. 

In this article, we establish a link between quantum mechanical molecular simulations and the T-matrix approach. The former provide the basic model for the latter. Their combination solves both the scalability issue of molecular simulations and the failure of macroscopic electromagnetism at the nano and meso scales. We first show how to build the T-matrix of a molecule using the dynamic polarizabilities obtained through quantum molecular simulations. The T-matrix of an object is equivalent to its scattering matrix and contains all the information about its electromagnetic response. Any electromagnetic property can be computed from it. We then provide an example of our approach where, starting from the simulation data of a chiral molecule obtained with the {\sc Turbomole} program package~\cite{tm1,tm2}, we compute chiro-optical properties for the single chiral molecule and a cross-like arrangement of four copies of it. For this purpose, we derive formulas to compute the Oriented Circular Dichroism (OCD) and rotationally averaged Circular Dichroism (CD) from the T-matrix. The expressions are not limited to the dipolar approximation: They are valid for arbitrarily high multipolar orders. They are hence useful for molecular ensembles whose sizes prevent the use of the dipolar approximation. We also provide formulas to quantify two properties that have been recently shown to be fundamentally relevant to chiral interactions: Electromagnetic duality and electromagnetic chirality. We describe both properties and comment on their technological significance. The link that we establish in this article allows to use existing algorithms for efficiently computing the joint electromagnetic response of a very large number of molecules. This approach can be used in the analysis and molecular-based design of discrete objects and materials for specific electromagnetic functions. Additionally, this work constitutes a necessary step towards the T-matrix based derivation of models and parameters for the constitutive relations of a bulk material from quantum mechanical simulations of its molecular constituents.

We start by connecting two different settings for describing light-matter interactions at the molecular level: The electric and magnetic polarizability tensor and the T-matrix. We assume that the molecule is immersed in an isotropic and homogeneous medium and roughly centered at position $\rrp$. The molecule is illuminated by an incident electromagnetic field, with which it interacts. A re-radiated(scattered) field results from the interaction. We restrict the following treatment to linear light-matter interactions which do not change the frequency of incident fields. That is, given the decomposition of the incident field into harmonic components with $\exp\left(-i\omega t\right)$ time dependency, each component will only cause the molecule to generate a scattered field of the same frequency $\omega$. We allow the electric permittivity $\epsilon$ and magnetic permeability $\mu$ of the surrounding medium to be frequency dependent $\left\{\epsilon^\omega,\mu^\omega\right\}$. We also assume that the dimensions of the molecule are much smaller than any of the wavelengths in the incident field. Under these conditions, there is a very common setting for modeling the light-matter interaction (see e.g. \cite[Eq. 6]{Autschbach2009} and \cite[Sec. II]{Sersic2011}): The scattered fields produced by the molecule at each frequency $\omega$ are due to the radiation of an electric dipole moment $\ed$ and a magnetic moment dipole $\md$ which are induced in the molecule by the incident electric and magnetic fields at the point $\rrp$: $\left\{\EE(\rrp),\HH(\rrp)\right\}$. The effect of the molecule is then completely characterized by a complex-valued 6$\times$6 matrix, which we here decompose in its four 3$\times$3 blocks using an obvious naming convention:
\begin{equation}
\label{eq:polarizability}
\begin{bmatrix}\ed\\\md\end{bmatrix}=
\begin{bmatrix}\dE & \dH\\\mE & \mH\end{bmatrix}
\begin{bmatrix}\EE(\rrp)\\\HH(\rrp)\end{bmatrix}.
\end{equation}
From now on, we choose the origin of the coordinate axis at $\rrp$ so that $\rrp=(0,0,0)$. SI units are assumed throughout, and the $\exp\left(-i\omega t\right)$ factors are suppressed.

A more general formalism for the study of light-matter interactions is the T-matrix setting. The T-matrix of an object relates the incident and scattered fields in the following way. First, a complete basis for free electromagnetic fields in the surrounding medium is chosen. The most common choice is the multipolar fields of well defined parity \cite{Mishchenko2016}. This is the most convenient option for our initial purposes. Each multipolar field is characterized by its frequency $\omega$, its total angular momentum squared $j(j+1)$ with $j=1,2,\ldots$, its angular momentum along one chosen axis $m=[-j\ldots j]$, its parity or electric/magnetic character, and, for the T-matrix formalism, its incident or scattered character. Scattered multipoles are purely outgoing, meet the radiation condition at infinity, and are singular at the origin, while incident multipoles are regular, i.e. do not have singularities, and are a of a mixed incoming and outgoing character. Using this basis, each frequency component of the incident electric and magnetic fields can be written as 
\begin{equation}
	\label{eq:EH}
	\begin{split}
\EE(\rr)=\sum_{j=1}^\infty\sum_{m=-j}^{m=j}\left[a^\omega_{jm}\NN_{jm}(\rr)+b^\omega_{jm}\MM_{jm}(\rr)\right],\\
iZ^\omega\HH(\rr)=\sum_{j=1}^\infty\sum_{m=-j}^{m=j}\left[b^\omega_{jm}\NN_{jm}(\rr)+a^\omega_{jm}\MM_{jm}(\rr)\right],
	\end{split}
\end{equation}
where $\{a^\omega_{jm},b^\omega_{jm}\}$ are complex coefficients, $\NN_{jm}(\rr)$ are regular electric multipolar fields, $\MM_{jm}(\rr)$ are regular magnetic multipolar fields, $Z^\omega=\sqrt{\frac{\mu^\omega}{\epsilon^\omega}}$, and the second line follows from the first by first using one of Maxwell's equations to show that $iZ_\omega\HH(\rr)=\nabla\times\EE(\rr)/k^\omega$, where $k^\omega=\omega\sqrt{\epsilon^\omega\mu^\omega}$, and the properties (\cite[Eq. 3.8]{Wheeler2010PHDTH}) $\nabla\times\MM_{jm}(\rr)=k^\omega\NN_{jm}(\rr)$ and $\nabla\times\NN_{jm}(\rr)=k^\omega\MM_{jm}(\rr)$. More explicit expressions for $\MM_{jm}(\rr)$ and $\NN_{jm}(\rr)$ are given in Eq. (\ref{eq:mps}) of App. \ref{app:connection}. An expansion similar to Eq. (\ref{eq:EH}), but featuring outgoing multipoles, holds for the scattered outgoing fields. We will denote by $c_{jm}^\omega$ the coefficients multiplying the scattered electric multipoles and by $d_{jm}^\omega$ the coefficients multiplying the magnetic ones. The T-matrix is then a matrix that relates the coefficients of the incident field with those of the scattered field:
\begin{equation}
	\label{eq:parity}
\begin{bmatrix}\vdots\\c_{jm}^\omega\\\vdots\\d_{jm}^\omega\\\vdots\end{bmatrix}=\T \begin{bmatrix}\vdots\\a_{jm}^\omega\\\vdots\\b_{jm}^\omega\\\vdots\end{bmatrix}.
\end{equation}
Since the integer index $j$ in Eq. (\ref{eq:EH}) ranges from $j=1$ to infinity, the dimensionality of the vectors and matrix in Eq. (\ref{eq:parity}) is infinite. Nevertheless, for any given object of finite size one can select a maximum multipolar order $j_{max}$ beyond which the interaction of the object with the electromagnetic field can be neglected. The choice of a $j_{max}$ makes the dimensions of the arrays in Eq. (\ref{eq:parity}) finite. For molecules illuminated with propagating beams of visible or UV light, taking $j_{max}=1$ (dipolar), or at most $j_{max}=2$ (quadrupolar) should suffice.

In order to connect Eqs. (\ref{eq:polarizability}) and (\ref{eq:parity}), we now apply to Eq. (\ref{eq:parity}) the restrictions contained in Eq. (\ref{eq:polarizability}). We first consider the restriction that the scattered field is produced by dipole moments induced in the molecule. For each frequency and each parity there are three dipolar fields corresponding to $j=1$ and $m\in[-1,0,1]$, for a total of 6 fields per frequency. Consequently, in the T-matrix setting, the fields scattered by the molecule need to be restricted to belong to the subspace expanded by these six dipolar fields. Equation (\ref{eq:polarizability}) also contains the restriction that the influence of the incident field is completely determined by its values at the origin. It turns out that this also restricts the incident fields to be only dipolar. It is easy to see from the actual expression of $\MM_{jm}(\rr)$ and $\NN_{jm}(\rr)$ at $\rr=(0,0,0)$ (see App. \ref{app:connection}) that:
\begin{equation}
	\label{eq:EHzero}
	\begin{split}
\EE(0)=\left[\sum_{m=-1}^{m=1}a^\omega_{1m}\NN_{1m}(0)\right],\\
iZ^\omega\HH(0)=\left[\sum_{m=-1}^{m=1}b^\omega_{1m}\NN_{1m}(0)\right].
	\end{split}
\end{equation}
Therefore, we can again write a 6$\times$6 linear relationship, this time in the T-matrix setting, which we also decompose into 3$\times$3 blocks:
\begin{equation}
\label{eq:tdipolar}
\cd=\begin{bmatrix}\Tnn&\Tnm\\\Tmn&\Tmm\end{bmatrix}\ab,
\end{equation}
where the subscripts labeling the blocks refer to the multipolar fields of the first line of Eq. (\ref{eq:EH}).

The models in Eqs. (\ref{eq:polarizability}) and (\ref{eq:tdipolar}) are physically equivalent. We show in App. \ref{app:connection} that the one-to-one relationship between the two 6$\times$6 matrices is:
{\small
\begin{eqnarray}
	\label{eq:connection}
&&\begin{bmatrix}\Tnn&\Tnm\\\Tmn&\Tmm\end{bmatrix}=\\
&&\frac{ic^\omega Z^\omega(k^\omega)^3}{6\pi}\begin{bmatrix}\C\left(\dE\right)\Cinv&\C\left(-i\dH/Z^\omega\right)\Cinv\\\C\left(i\mE/c^\omega\right)\Cinv&\C\left(\mH/(c^\omega Z^\omega)\right)\Cinv\end{bmatrix},\nonumber
\end{eqnarray}
}
where $c^\omega=1/\sqrt{\epsilon^\omega\mu^\omega}$ is the frequency dependent speed of light in the surrounding medium, and $\C$ is the 3$\times$3 unitary change of basis matrix that goes from the Cartesian to the spherical basis [see Eq. (\ref{eq:cartsph})]. This change of basis is needed in the most common case where Eq. (\ref{eq:polarizability}) is expressed in the Cartesian basis. That is: $\ed=[p^\omega_x,p^\omega_y,p^\omega_z]$, $\EE=[E^\omega_x,E^\omega_y,E^\omega_z]$, etc $\ldots$ . On the other hand, Eq. (\ref{eq:tdipolar}) is an expression related to the spherical basis.

Equation (\ref{eq:connection}) connects the two settings and allows to build the T-matrix of the molecule to dipolar order using quantum chemical molecular simulations to compute the tensors $\dE$, $\dH$, $\mE$ and $\mH$. 

A comment on the sources of inaccuracy of the dipolar approximation is now pertinent. Theoretically, the only approximation is having neglected quadrupolar and higher order terms. In practice, there is at least another source of inaccuracy in the dipolar terms themselves. As in many other cases, the dipole moments in molecular simulations are usually computed using long wavelength approximations of the exact expressions \cite{FerCor2015b,Alaee2016b}.

The T-matrix setting is very useful for going beyond the response of single molecules to that of ensembles of molecules and even to the response of bulk materials. One of the crucial features of the T-matrix setting is that, given the T-matrices of several objects, the calculation of the T-matrix of an arbitrary arrangement of them can be performed efficiently. The technique (see e.g. \cite[Sec. 4]{Mischchenko1996}) allows to rigorously account for the inter-particle electromagnetic coupling in the calculation of the response of the composite object. The key to compute the inter-particle couplings are the translation theorems of vector spherical harmonics. They allow to ``translate'' the multipolar radiation of one object onto the location of a second object. The field re-radiated by the second object due to the radiation from the first one is then computed using the T-matrix of the second object. After considering all the objects and mutual interactions, a self-consistent set of equations is obtained whose solution gives the T-matrix of the ensemble. The only limitation is that there cannot be currents flowing from one object to the other. In our context, it means that there cannot be charges flowing between any two molecules of the ensemble. This restriction would not be met if there exist covalent bounds between molecules. Provided that the restriction is met, the T-matrix route to the calculation of the response of an ensemble of molecules is much more efficient than the direct molecular simulation of the ensemble. While state-of-the-art T-matrix codes can handle ensembles of tens of thousands of individual objects \cite{Egel2017}, molecular simulations of the same size are unfeasible. It is also noteworthy to mention that the dipolar approximation taken on each individual molecule does not preclude the computation of the T-matrix of the ensemble to any desired multipolar order, only limited by the available numerical resources. Methods based on the T-matrix approach allow to obtain the response of finite or infinite, periodic or aperiodic 2D and 3D arrays of molecules from the T-matrices of the individual molecules (see \cite{Xu2013,Xu2014} and the references in \cite[Sec. 2.6]{Mischchenko1996}). Similarly, the computation of the joint responses of large numbers of randomly arranged molecules is also possible \cite[Sec. 2.7]{Mischchenko1996}. Equation (\ref{eq:connection}) allows the use of all these existing algorithms in molecular based material design.

When the maximum multipolar order in the T-matrix is large enough so that the influence of the omitted higher orders can be neglected, the T-matrix is a complete description of the electromagnetic response of the object. This means that any electromagnetic property of the object can be computed from it: Scattering cross-sections, absorption cross-sections, etc ... . We will now focus on the calculation of chiro-optical properties of individual molecules and molecular ensembles using the T-matrix. We start with a change of basis that is very useful for this purpose. The following change of basis:
\begin{equation}
	\label{eq:change}
	\begin{split}
		\Ajmp(\rr)&=\frac{\NN_{jm}(\rr)+\MM(\rr)_{jm}}{\sqrt{2}},\\
		\Ajmm(\rr)&=\frac{\NN_{jm}(\rr)-\MM_{jm}(\rr)}{\sqrt{2}}, 
	\end{split}
\end{equation}
goes from multipolar fields of well defined parity to multipolar fields of well defined helicity. That is, spherical waves whose plane wave decompositions contain a single polarization handedness. In the helicity basis, Eq. (\ref{eq:parity}) reads: 
\begin{equation}
	\label{eq:helicity}
\begin{bmatrix}\vdots\\\rho_{jm+}^\omega\\\vdots\\\rho_{jm-}^\omega\\\vdots\end{bmatrix}=\begin{bmatrix}\Tpp&\Tpmi\\\Tmip&\Tmimi \end{bmatrix}\begin{bmatrix}\vdots\\\mu_{jm+}^\omega\\\vdots\\\mu_{jm-}^\omega\\\vdots\end{bmatrix}.
\end{equation}
Appendix \ref{app:pth} contains the expressions of the elements in Eq. (\ref{eq:helicity}) as a function of elements in Eq. (\ref{eq:parity}).

Given its T-matrix, the absorption $a_{\pm}^\omega(\vhat)$ of any individual or composite object upon illumination with a circularly polarized plane wave with momentum direction $\vhat$ and of either +1 or -1 helicity, i.e. either left or right hand polarized, can be written in this formulation (see App. \ref{app:OCD}) as
{\small
\begin{equation}
	\label{eq:abs}
	\begin{split}
		a_+^\omega(\vhat)&=\\-\frac{1}{2}{\mupv}^\dagger & \left({\Tpp}^\dagger+\Tpp+2{\Tpp}^\dagger\Tpp+2{\Tmip}^\dagger \Tmip\right)\mupv\\
		&={\mupv}^\dagger\matr{A}_+^\omega\mupv,\\
		a_-^\omega(\vhat)&=\\-\frac{1}{2}{\mumv}^\dagger & \left({\Tmimi}^\dagger+\Tmimi+2{\Tmimi}^\dagger\Tmimi+2{\Tpmi}^\dagger \Tpmi\right)\mumv\\
		&={\mumv}^\dagger\matr{A}_-^\omega\mumv.\\
	\end{split}
\end{equation}
}
where $\dagger$ denotes conjugate transposition. The vectors $\mupv$ and $\mumv$ contain the complex expansion coefficients in the multipolar helicity basis of a plane wave with momentum aligned along $\vhat$ and either +1 or -1 helicity (see App. \ref{app:OCD}). It is obvious that the oriented circular dichroism [OCD$^\omega(\vhat)$] of the object is just $a_+^\omega(\vhat)-a_-^\omega(\vhat)$. 

With respect to the rotationally averaged CD (CD$^\omega$), App. \ref{app:CD} contains the proofs that:
\begin{equation}
	\label{eq:CDexplicittext}
	\text{CD}^\omega=4\pi\Tr{\matr{A}_+^\omega-\matr{A}_-^\omega},
\end{equation}
and that, in the common units of [liter/mol/cm]:
\begin{equation}
	\label{eq:CDbarexplicittext}
	\overline{\text{CD}}^\omega=\frac{10N_A}{\log(10)}\frac{4\pi}{{(k^\omega)}^2}\Tr{\matr{A}^\omega_+-\matr{A}^\omega_-}.
\end{equation}
where $\Tr{\matr{F}}$ denotes the trace of the matrix $\matr{F}$, and $N_A$ is Avogadro's number.

In order to illustrate the discussion, we provide an example based on a sodium salt of D-camphoric acid. In this example, we neglect $\mH$ and compute the tensors $\dE$, $\dH$ and $\mE$ using damped response theory, also known as complex polarization propagator theory, at the TD-DFT level~\cite{Jiemchooroj2007,Cukras2016}. The tensor $\dE$ is the damped electric-dipole--electric dipole linear response function and  $\dH$ and $\mE$ are obtained from the damped mixed electric-dipole--magnetic dipole linear response function. We first compute the damped electric-dipole--electric dipole and mixed electric-dipole--magnetic dipole linear response functions for a single molecule. These computations were performed at the Hartree--Fock level in the def2-TZVPD basis set \cite{Rappoport2010} of Gaussian atomic orbitals using the {\sc Turbomole} program package~\cite{tm1,tm2}. The linear response functions were obtained using sum-over-states expressions including the electric and magnetic transition dipoles (in the length representation) of all 42368 singlet excited time-dependent Hartree--Fock (TD-HF) states of the molecule. The damping parameter used corresponds to a Lorentzian line shape with full width at half maximum (FWHM) of 0.5 eV.

\begin{figure}[h!]
	\includegraphics[width=0.6\linewidth]{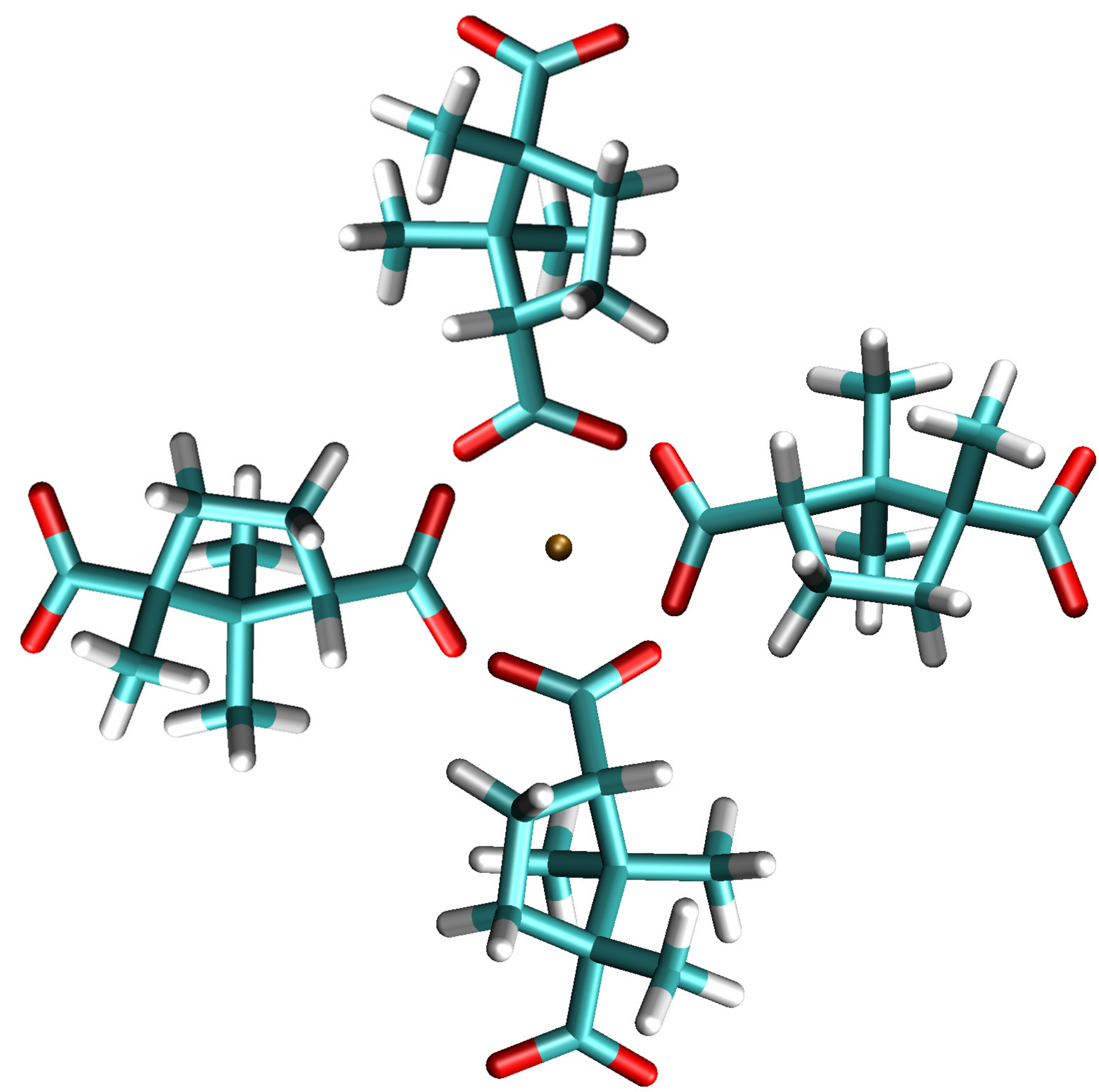}
	\caption{Cross-like arrangement of four identical D-camphoric acid molecules.\label{fig:cross}} 
\end{figure}

\begin{figure}[h!]
	\includegraphics[width=\linewidth]{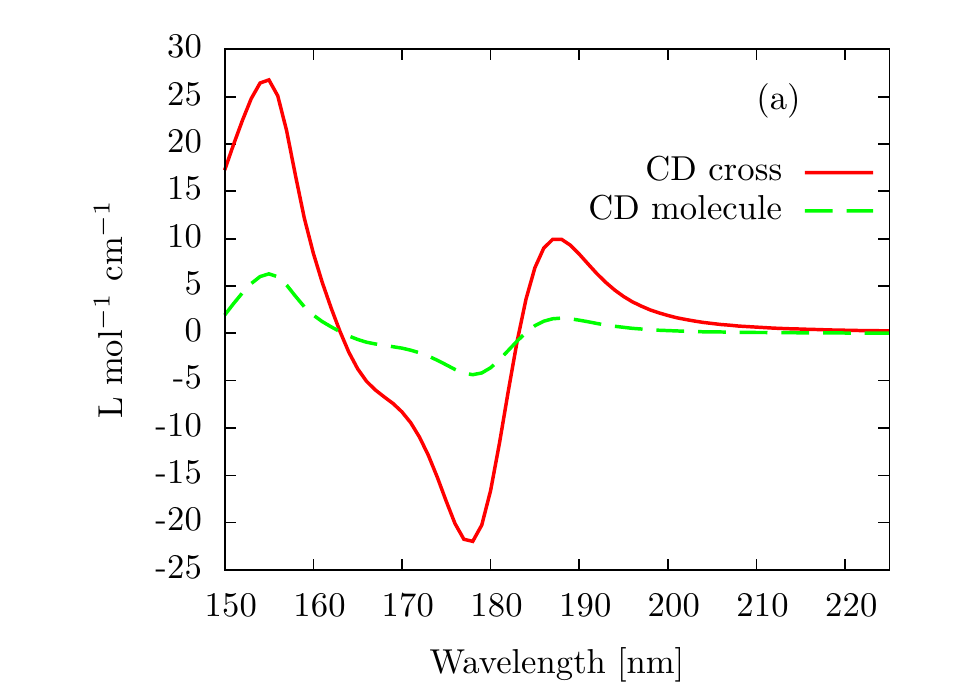}\\
	\includegraphics[width=\linewidth]{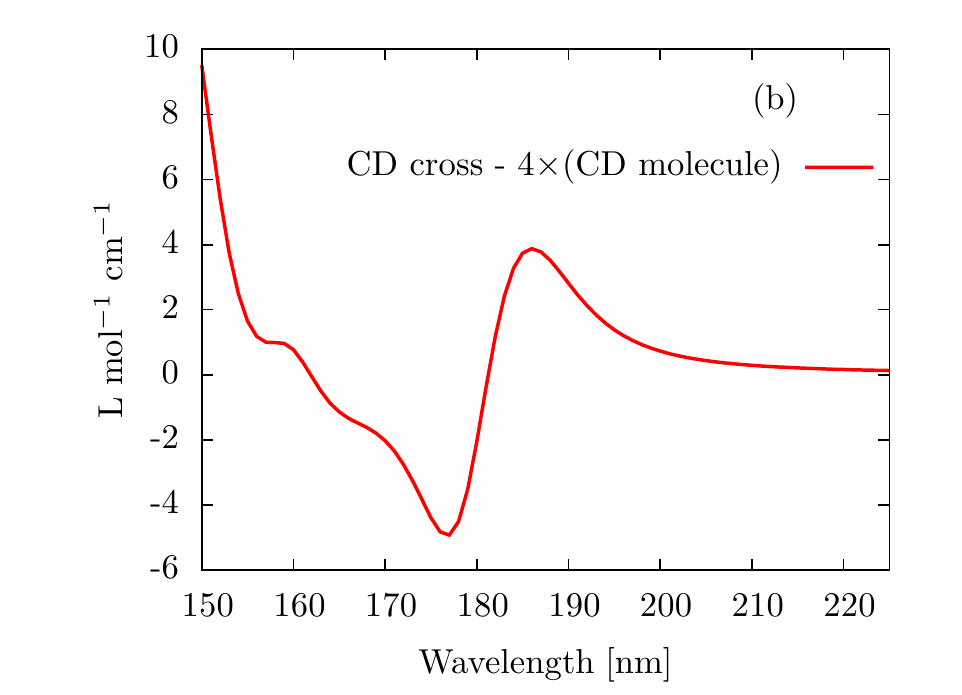}\\
	\caption{(a) Rotationally averaged CD of a single molecule (green dashed line) and the cross-like structure shown in Fig. \ref{fig:cross} (solid red line). (b) Difference between the rotationally averaged CD of the cross-like structure and four times the one of a single molecule. \label{fig:cd}} 
\end{figure}

The molecular simulation data can then be used to build the T-matrix of the molecule using Eq. (\ref{eq:connection}). Then, the T-matrix of the cross-like arrangement of four copies of it shown in Fig. \ref{fig:cross} can be obtained by existing algorithms (see e.g. \cite{Mischchenko1996}). We set $j_{max}=2$. It should be noted that such cross-like structure can serve as a model for the unit cell of the layers that compose chiral SURMOFs \cite{Gu2014}. The T-matrix of the cross is hence the main ingredient for computations of the response of the SURMOFs. 
With respect to chiro-optical properties, Fig. \ref{fig:cd}(a) shows the averaged CD for both the molecule and the cross-like arrangement computed with Eq. (\ref{eq:CDbarexplicittext}). We have verified that, for the single molecule, the results exactly match those obtained with established methodologies. Figure \ref{fig:cd}(b) shows the difference between the averaged CD of the cross and four times that of the single molecule. This difference is due to electromagnetic inter-molecule coupling. 

Recently, two concepts have been shown to be of fundamental importance in chiral interactions: Electromagnetic duality symmetry and electromagnetic chirality. The T-matrix is a very convenient object to quantitatively study them. We start with duality symmetry.

A system that responds in the same way to electric and magnetic fields is said to have electromagnetic duality symmetry, or, for short, to be dual. For a dual-symmetric system, any incident and scattered fields that are a solution of the light-matter interaction problem produce a new solution by transforming all the fields as:
\begin{equation}
		\label{eq:duality}
		\begin{split}
				\EE_\theta&=\EE\cos\theta  - Z^\omega\HH\sin\theta,\\
		Z^\omega\HH_\theta&=\EE\sin\theta + Z^\omega\HH\cos\theta,
		\end{split}
\end{equation}
for any $\theta$. In much the same way that a translationally invariant system prevents the coupling of plane waves with different momenta during the light-matter interaction, a dual system prevents the coupling of the two helicity components of the field. Among other phenomena, electromagnetic duality is important in optical activity. Recent work \cite{FerCor2012c} shows that, given a system and a pair of incident and scattered plane wave directions, the non-mixing of the two helicities by the system is a necessary condition for optical rotation. This condition is in addition to the lack of mirror symmetry across the plane defined by the incident and scattered direction vectors. For the particular case of forward transmission through a random medium like a solution of chiral molecules, the non-mixing is achieved by means different than duality symmetry but, when the measurements are performed at an angle, there is mixing of the two helicities and the polarization rotation angle depends on the input polarization \cite{Vidal2015}. The off-axis behavior is hence qualitatively different from the forward direction where the rotation angle is an additive constant independent of the input polarization angle. For optical activity in general incident and scattered directions, a dual system is required \cite{FerCor2015}. The degree with which the optical rotation angles vary with the input polarization depends on the amount of mixing between the two helicities. Equation (2) in \cite{FerCor2015} defines a convenient measure of helicity mixing, i.e. duality breaking. The measure produces a number between 0 and 1, with 0 corresponding to perfect duality symmetry. It can be computed as:
\begin{equation}
	\label{eq:ddash}
		\cancel{D}^\omega=\frac{\Tr{{\Tpmi}^\dagger\Tpmi}+\Tr{{\Tmip}^\dagger\Tmip}}{\Tr{{\T}^\dagger{\T}}}. 
\end{equation}
Dual systems are also needed for other technologically important concepts like zero-backscattering, metamaterials for transformation optics, Huygens wave-front control, and maximal electromagnetic chirality (explained below). Unfortunately, naturally dual bulk materials do not exist, and their artificial fashioning has only been achieved at MHz frequencies \cite{sengupta2000,Schubring2004}. The possibility of achieving materials with high duality by molecular design and assembly is therefore of great technological interest. 

With respect to chirality, while the definition of when an object is chiral is simple enough, it hides significant problems that arise when attempting to measure chirality \cite{Fowler2005}. It has been shown that a scalar measure of chirality allowing to rank general objects and/or to establish what a maximally chiral object is in an unambiguous way does not exist \cite{Buda1992,Petitjean2003,Rassat2004}. Recently, these problems have been solved by the definition of the electromagnetic chirality (em-chirality) of an object \cite{FerCor2016}, which is based on interaction instead of geometry. It can be stated in the following way: An {\em electromagnetically chiral object} is one for which all the information obtained from experiments using a fixed incident helicity {\em cannot} be obtained using the opposite one. The electromagnetic chirality of a given object has an upper bound. In a monochromatic setting, the upper bound of the em-chirality of an object is equal to $\sqrt{C^\omega}=\sqrt{\Tr{{\T}^\dagger \T}}$, which can be seen as a measure of how much does the object interact with the electromagnetic field. It turns out that objects which attain the bound are transparent to all the fields of one helicity. This makes them ideal for applications like helicity dependent photon routing (see \cite[Figs. 4]{FerCor2016}), among others. For reciprocal objects, maximal em-chirality implies duality symmetry. The normalized measure of em-chirality is a number between 0 and 1 which, in the context of this article, can be computed using the singular value decomposition of the T-matrix blocks in the helicity basis:
\begin{equation}
	\label{eq:chi}
\chi^\omega=\frac{\sqrt{\left\|\begin{bmatrix}\sigma(\Tpp)\\\sigma(\Tmip)\end{bmatrix}-\begin{bmatrix}\sigma(\Tmimi)\\\sigma(\Tpmi)\end{bmatrix}\right\|^2}}{\sqrt{C^\omega}},
\end{equation}
where $\sigma(A)$ denotes the column vector of non-increasingly ordered singular values of matrix $A$. We note that $C^\omega$ can also be written as
\begin{equation}
	\begin{split}
			C^\omega&=\sigma(\Tpp)^T\sigma(\Tpp)+\sigma(\Tmip)^T\sigma(\Tmip)\\
				 &+\sigma(\Tpmi)^T\sigma(\Tpmi)+\sigma(\Tmimi)^T\sigma(\Tmimi),
	\end{split}
\end{equation}
where $^T$ denotes transposition, which makes $\chi^\omega$ in Eq. (\ref{eq:chi}) a unitless quantity.
\begin{figure}[h!]
	\includegraphics[width=\linewidth]{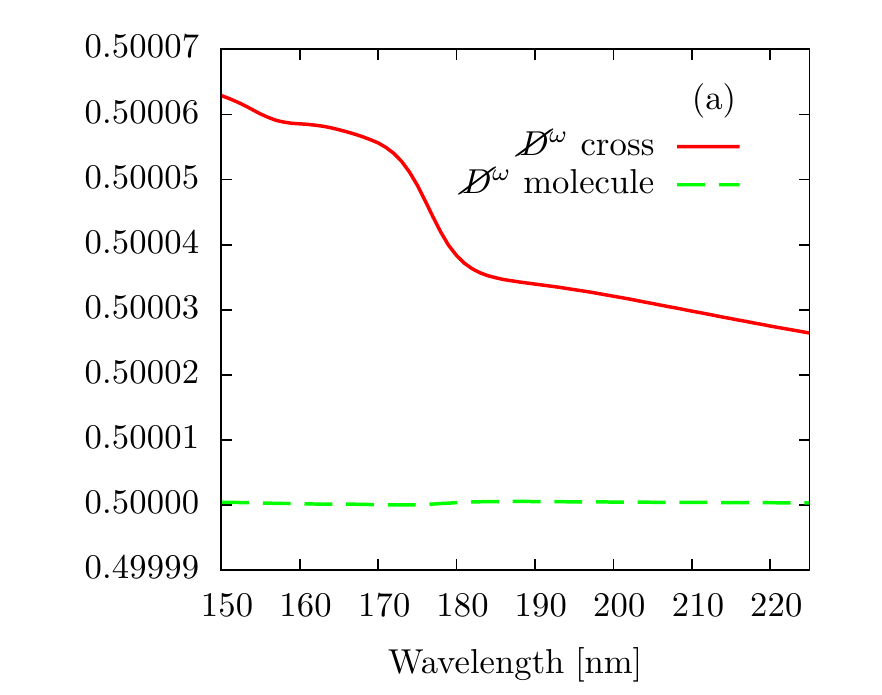}
	\includegraphics[width=\linewidth]{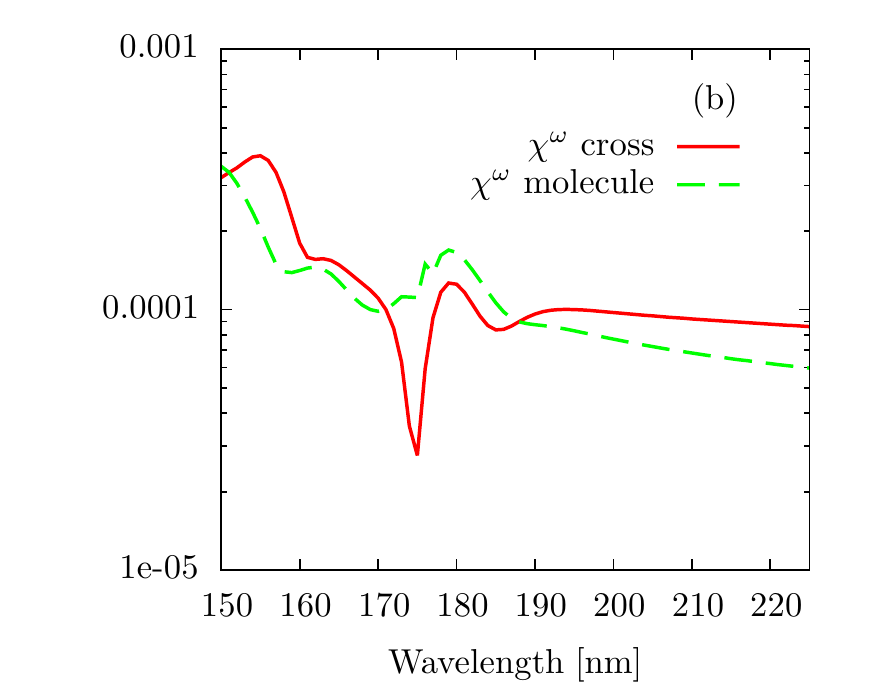}\\
	\caption{(a) Frequency dependent duality breaking ($\cancel{D}^\omega$) and, (b) normalized electromagnetic chirality ($\chi^\omega$) of both the individual molecule and the cross-like structure shown in Fig. \ref{fig:cross}. Both quantities are unitless [see Eqs. (\ref{eq:ddash}) and (\ref{eq:chi})]. \label{fig:new}} 
\end{figure}
As with duality, achieving materials with high em-chirality by molecular design and assembly is of great technological interest. Furthermore, the measure of electromagnetic chirality opens up a path to tackle the quantitative understanding of how the chirality of a composite object builds up from the chirality of its components and their arrangement. Chirality is present across different spatial scales that span several orders of magnitude from high energy physics to biology, but the mechanisms by which chirality spans across spatial scales are not clear, as written in a recent review \cite{Morrow2017}, ``how chirality at one length scale can be translated to asymmetry at a different scale is largely not well understood''. The impossibility of comparing the chiralities of two different systems in an unambiguous way is now solved by the measure of em-chirality, and a quantitative study of how em-chirality is transmitted across different spatial scales becomes possible. 

Going back to our example, Fig. \ref{fig:new}(a) shows the duality breaking $\cancel{D}$ of both the single molecule and the cross. The high values of duality breaking $\approx 0.5$ are consistent with objects whose electric-electric response $\dE$ is much larger than the magnetic-magnetic response $\mH$. Perfect duality symmetry at the dipolar level is met iff \cite[Eq. 29]{FerCor2013}: $\dE=\epsilon^\omega\mH$ and $\dH=-\mu^\omega\mE$. Figure \ref{fig:new}(b) shows the normalized em-chiralities, which are rather low for both the individual molecule and the ensemble.

While the results in Fig. \ref{fig:new} are far from the technologically desired ones, the example shows the usefulness of the framework put forward in this article to i) compute T-matrices of individual molecules using molecular simulations, ii) use the T-matrices of individual molecules to obtain the T-matrices of molecular ensembles, and iii) compute chiro-optical properties using the T-matrices. 

In conclusion, it is now possible to efficiently compute the response of a very large number of molecules. This can be applied in the analysis and molecular-based design of discrete objects and materials for specific electromagnetic functions.

\begin{acknowledgments}
I.F.-C.\ and C.R.\ gratefully acknowledge support by the VIRTMAT project at KIT.
\end{acknowledgments}
 
\appendix
\section{Connection\label{app:connection}}
In this appendix we will prove Eq. (\ref{eq:connection}), which connects the T-matrix setting
\begin{equation}
\label{eq:tdipolarApp}
\cd=\begin{bmatrix}\Tnn&\Tnm\\\Tmn&\Tmm\end{bmatrix}\ab,
\end{equation}
to the polarizability tensor setting
\begin{equation}
\label{eq:polarizabilityApp}
\begin{bmatrix}\ed\\\md\end{bmatrix}=
\begin{bmatrix}\dE & \dH\\\mE & \mH\end{bmatrix}
\begin{bmatrix}\EE(\mathbf{0})\\\HH(\mathbf{0})\end{bmatrix},
\end{equation}
and allows to build the T-matrix of a molecule to dipolar order using data obtained from quantum chemistry molecular simulations.

We will start from Eq. (\ref{eq:tdipolarApp}) and transform it towards Eq. (\ref{eq:polarizabilityApp}). We begin by connecting the $\{a^\omega_{1m},b^\omega_{1m}\}$ to $\{\EE(\mathbf{0}),\HH(\mathbf{0})\}$. The bottom line is that the $\{a^\omega_{1m},b^\omega_{1m}\}$ are essentially the coordinates of $\{\EE(\mathbf{0}),\HH(\mathbf{0})\}$ in the spherical vector basis.

Let us start by writing expressions for the multipolar fields of well defined parity $\MM_{jm}(\rr)$ and $\NN_{jm}(\rr)$:
\begin{equation}
	\label{eq:mps}
	\begin{split}
		\MM_{jm}(\rr)=&j_j(k^\omega r)\TT_{jjm}(\rhat),\\
		\NN_{jm}(\rr)=&\frac{\nabla\times\MM_{jm}(\rr)}{k^\omega}=-i\sqrt{\frac{j}{2j+1}}j_{j+1}(k^\omega r)\TT_{jj+1m}(\rhat)+\\&i\sqrt{\frac{j+1}{2j+1}}j_{j-1}(k^\omega r)\TT_{jj-1m}(\rhat).
	\end{split}
\end{equation}
where $k^\omega=\omega\sqrt{\epsilon^\omega\mu^\omega}$ is the frequency dependent wavenumber, $r=|\rr|$, $\rhat=\rr/|\rr|$, $j_l(\cdot)$ are spherical Bessel functions, and $\TT_{jlm}(\rhat)$ are the vector spherical harmonics as defined in \cite[Eq. 16.88]{Arfken2012}. Importantly, the spherical Bessel functions contain all the radial dependence of the multipolar fields, and the $\TT_{jlm}(\rhat)$ all their angular dependence. The second equality in the second line of Eq. (\ref{eq:mps}) follows from \cite[Eq. 16.100]{Arfken2012} and the relationships between spherical Bessel functions in \cite[p. 172]{Tung1985}.

At the origin ($\rr=\mathbf{0}$) the spherical Bessel functions $j_l(\cdot)$ are all zero except when $l=0$: $j_0(0)=1$. It follows that from all the multipolar fields in Eq. (\ref{eq:mps}), only the ones containing $j_0(0)$ are non-zero at the origin. Since $j=1,2,\ldots$, only electric dipolar fields
\begin{equation}
	\label{eq:nnt}
	\NN_{1m}(\mathbf{0})=i\sqrt{\frac{2}{3}}\TT_{10m}(\rhat),
\end{equation}
where $m=[-1,0,1]$ are non-zero at the origin. Using \cite[Eq. 16.88]{Arfken2012} and a table of Clebsch-Gordan coefficients we obtain
\begin{equation}
	\label{eq:tt}
		\TT_{10-1}(\rhat)=\frac{\ehatm}{\sqrt{4\pi}},\ \TT_{100}(\rhat)=\frac{\ehatz}{\sqrt{4\pi}},\ \TT_{101}(\rhat)=\frac{\ehatp}{\sqrt{4\pi}}.
\end{equation}
where $\{\ehatm,\ehatz,\ehatp\}$ are the spherical vector basis. We write a vector $\wvec$ in the spherical vector basis as:
\begin{equation}
	\wvec=w_{-1}\ehatm+w_0\ehatz+w_{1}\ehatp,
\end{equation}
with
\begin{equation}
	\label{eq:bvec}
	\ehatm=\frac{\xhat-i\yhat}{\sqrt{2}},\ 	\ehatp=-\frac{\xhat+i\yhat}{\sqrt{2}},\ \ehatz=\zhat.
\end{equation}
This choice of basis induces the following relationship between the Cartesian and spherical coordinates of $\wvec$ in the spherical and Cartesian basis:
\begin{equation}
	\label{eq:cartsph}
\begin{bmatrix}w_{-1}\\w_0\\w_{1}\end{bmatrix}=
	\begin{bmatrix}
		\frac{1}{\sqrt{2}}&\frac{i}{\sqrt{2}}&0\\
		0&0&1\\
		\frac{-1}{\sqrt{2}}&\frac{i}{\sqrt{2}}&0\\
	\end{bmatrix}.
\end{equation}
We can now use Eqs. (\ref{eq:nnt}) and (\ref{eq:tt}) to write a more explicit version of Eq. (\ref{eq:EHzero}):
\begin{equation}
	\label{eq:EHzeroexp}
	\begin{split}
		\EE(0)=i\sqrt{\frac{2}{12\pi}}\left[\sum_{m=-1}^{m=1}a^\omega_{1m}\ehatmm\right],\\
			iZ^\omega\HH(0)=i\sqrt{\frac{2}{12\pi}}\left[\sum_{m=-1}^{m=1}b^\omega_{1m}\ehatmm\right],
	\end{split}
\end{equation}
which says that the $a^\omega_{1m}$ and $b^\omega_{1m}$ are respectively proportional to the coordinates of the incident electric and magnetic fields at the origin in the {\em spherical} vector basis coordinates.

We now turn to the left hand side of Eq. (\ref{eq:tdipolarApp}). We need to relate the electric and magnetic dipole moments $\ed$ and $\md$ to the multipolar coefficients $c^\omega_{1m}$ and $d^\omega_{1m}$ of their radiated fields. This task can be achieved using expressions from \cite[Chap. 9]{Jackson1998}. The idea is to equate two different expressions of the magnetic(electric) field radiated by an electric(magnetic) dipole moment. One of the expressions involves $\ed$($\md$) in \cite[Eq. (9.19)]{Jackson1998}(\cite[Eq. (9.36)]{Jackson1998}), and the other the multipolar coefficients of the field \cite[Eq. (9.149)]{Jackson1998}(far field limit of \cite[Eq. (9.122)]{Jackson1998} which can be obtained using \cite[Eq. (9.89)]{Jackson1998}). One then uses an expression of the cross product in spherical coordinates
\begin{equation}
	\sum_{m=-1}^{m=1}w_m\mathbf{X}_{1m}(\rhat)=i\sqrt{\frac{3}{8\pi}}{\mathbf{w}}\times\rhat,
\end{equation}
with the definition of $\mathbf{X}_{1m}(\rhat)$ in \cite[Eq. (9.119)]{Jackson1998}, and the scale factor differences between the coefficients in the multipole expansions that we are using [Eq. (\ref{eq:EH})] and those in \cite[Eq. (9.122)]{Jackson1998} to conclude that:
\begin{equation}
\label{eq:hard} 
\begin{bmatrix}c^\omega_{1-1}\\c^\omega_{10}\\c^\omega_{11}\end{bmatrix}=\frac{c^\omega Z^\omega (k^\omega)^3}{\sqrt{6\pi}}\ed,\ \begin{bmatrix}d^\omega_{1-1}\\d^\omega_{10}\\d^\omega_{11}\end{bmatrix}=i\frac{ Z^\omega (k^\omega)^3}{\sqrt{6\pi}}\md,
\end{equation}
where $c^\omega=1/\sqrt{\epsilon^\omega\mu^\omega}$. 

We can now perform the connection between Eq. (\ref{eq:tdipolarApp}) and (\ref{eq:polarizabilityApp}). First we use Eq. (\ref{eq:hard}) on the left hand side of Eq. (\ref{eq:tdipolarApp}), and Eq. (\ref{eq:EHzeroexp}) on the vector of its right hand side:
\begin{equation}
\frac{c^\omega Z^\omega (k^\omega)^3}{\sqrt{6\pi}}\begin{bmatrix}\ed\\\frac{i\md}{c^\omega}\end{bmatrix}=\begin{bmatrix}\Tnn&\Tnm\\\Tmn&\Tmm\end{bmatrix}\begin{bmatrix}\EE(0)\\iZ^\omega\HH(0)\end{bmatrix}(-i)\sqrt{\frac{12\pi}{2}}
\end{equation}
Re-arranging the scalar factors we obtain
\begin{equation}
\begin{bmatrix}\ed\\\frac{i\md}{c^\omega}\end{bmatrix}=\begin{bmatrix}\Tnn&\Tnm\\\Tmn&\Tmm\end{bmatrix}\begin{bmatrix}\EE(0)\\iZ^\omega\HH(0)\end{bmatrix}\frac{(-i)6\pi}{c^\omega Z^\omega (k^\omega)^3},
\end{equation}
which, after modifying the 3$\times$3 matrices to obtain the desired input and output vectors produces
\begin{equation}
\label{eq:jahisom}
\begin{bmatrix}\ed\\\md\end{bmatrix}=\begin{bmatrix}\Tnn&iZ^\omega\Tnm\\-ic^\omega\Tmn&c^\omega Z^\omega\Tmm\end{bmatrix}\begin{bmatrix}\EE(0)\\\HH(0)\end{bmatrix}\frac{(-i)6\pi}{c^\omega Z^\omega (k^\omega)^3}.
\end{equation}
And it follows form comparing Eq. (\ref{eq:polarizabilityApp}) with Eq. (\ref{eq:jahisom}) that 
\begin{equation}
	\label{eq:final}
\begin{bmatrix}\dE & \dH\\\mE & \mH\end{bmatrix}=\frac{(-i)6\pi}{c^\omega Z^\omega (k^\omega)^3}\begin{bmatrix}\Tnn&iZ^\omega\Tnm\\-ic^\omega\Tmn&c^\omega Z^\omega\Tmm\end{bmatrix}.
\end{equation}
This result leads directly to Eq. (\ref{eq:connection}) by recalling that in the typical case where the polarizabilities are available in the Cartesian basis, we must change the basis to spherical. The change of basis matrix $\C$ in Eq. (\ref{eq:connection}) is the one in Eq. (\ref{eq:cartsph}).

\section{Absorption\label{app:absorption}}
The absorption of an object upon a particular illumination is best analyzed using the S-matrix. Given the T-matrix, the S-matrix or scattering matrix can be computed as:
\begin{equation}
		\label{eq:ST}
		\matr{S}^\omega=\matr{I}+2\matr{T}^\omega,
\end{equation}
where $I$ is the identity matrix. Notwithstanding the simple numerical relationship in Eq. (\ref{eq:ST}), there is an important physical difference between the two matrices. While the T-matrix relates incident and scattered fields, the S-matrix relates total incoming and outgoing fields. The total incoming(outgoing) fields are the total fields before(after) the interaction. The difference is that an incident field has a mixed incoming and outgoing character and exists before and after the interaction. Its incoming part is the total incoming field, its outgoing part plus the scattered field equals the total outgoing field. The fact that the S-matrix connects total fields allows for the simple following derivation of the absorption. 
\subsection{Oriented CD\label{app:OCD}}
Having assumed that the light-matter interaction does not change the frequency of the fields, we can just focus on a single frequency component. Let us consider an incoming field represented by its coordinates in the helicity basis $\alphav$. Its norm squared is ${\alphav}^{\dagger}\alphav$. The outgoing field is $\SSS\alphav$, with norm squared ${\alphav}^{\dagger}{\SSS}^\dagger\SSS\alphav$. The absorption must hence be the difference
\begin{equation}
		\label{eq:diff}
		{\alphav}^{\dagger}\alphav-{\alphav}^{\dagger}{\SSS}^\dagger\SSS\alphav.
\end{equation}
After using the block decomposition of the T-matrix in Eq. (\ref{eq:helicity}), Eq. (\ref{eq:diff}) becomes
{\small
\begin{equation}
		\label{eq:ss}
{\alphav}^{\dagger}\alphav-{\alphav}^{\dagger}\left(\matr{I}+2\begin{bmatrix}\Tpp&\Tpmi\\\Tmip&\Tmimi \end{bmatrix}\right)^\dagger\left(\matr{I}+2\begin{bmatrix}\Tpp&\Tpmi\\\Tmip&\Tmimi \end{bmatrix}\right)\alphav.
\end{equation}
}
Particularizing $\alphav$ to a plane wave of a given helicity with momentum direction $\vhat$ quickly leads to Eq. (\ref{eq:abs}). Let us do it for an incoming plane wave of positive helicity, which has zero projection on the negative helicity multipoles $\Ajmm(\rr)$ [Eq. (\ref{eq:change})], so half of the vector is filled with zeros:
\begin{equation}
\alphawithzeros.
\end{equation}
Then Eq. (\ref{eq:ss}) can be written as:
\begin{equation}
		\begin{split}
		&{\alphav_+(\vhat)}^{\dagger}\alphav_+(\vhat)-\\
		&\left({\alphawithzeros}^{\dagger}+2\begin{bmatrix}{\alphav_+(\vhat)}^{\dagger}{\Tpp}^\dagger&{\alphav_+(\vhat)}^{\dagger}{\Tmip}^\dagger\end{bmatrix}\right)\times\\
		&\left(\alphawithzeros+2\begin{bmatrix}\Tpp\alpha_+(\vhat)\\\Tmip\alpha_+(\vhat)\end{bmatrix}\right),
		\end{split}
\end{equation}
which readily results into the first line of Eq. (\ref{eq:abs}) after considering the following point. The vectors of coefficients $\mu_\pm(\vhat)$ in Eq. (\ref{eq:abs}) represent {\em incident} plane waves, while the vectors of coefficients $\alpha_\pm(\vhat)$ represent the {\em incoming} part of the incident plane waves. The numerical relation 
\begin{equation}
	\alpha_\pm(\vhat)=\frac{1}{2}\mu_\pm(\vhat),
\end{equation}
can be deduced from the expansion of an incident plane wave into regular multipoles, featuring spherical Bessel functions, which can be written as the following sum of incoming and outgoing spherical Hankel functions: $j_l(\cdot)=[h^1_l(\cdot)+h^2_l(\cdot)]/2$.
\subsection{Rotationally averaged CD\label{app:CD}}
We now address the rotationally averaged CD, that is, the differential absorption averaged over all possible spatial directions of an incident plane wave:
\begin{equation}
	\label{eq:cd1}
	\text{CD}^\omega=\int d\vhat\left[\mupv^\dagger\matr{A}^\omega_+\mupv-\mumv^\dagger\matr{A}^\omega_-\mumv\right],
\end{equation}
where $\matr{A}_\pm^\omega$ are defined in Eq. (\ref{eq:abs}). As per \cite[Eq. (8.4-6)]{Tung1985}, each $\mupv$ and $\mumv$ can be obtained by a corresponding rotation of a reference vector representing a plane wave whose momentum is aligned with the $\zhat$ direction: $\vec{\mu}_{\pm}(\vhat)=\matr{R}(\vhat)\vec{\mu}_{\pm}(\zhat)$. This allows us to write Eq. (\ref{eq:cd1}) as
\begin{equation}
	\label{eq:CD}
		\text{CD}^\omega=\mupvz^\dagger\RAp\mupvz-\mumvz^\dagger\RAm\mumvz,
\end{equation}
where $\text{Ro}\left\{\matr{F}\right\}=\int d\vhat \matr{R}(\vhat)^\dagger \matr{F}\matr{R}(\vhat)$ is the rotational average of matrix $\matr{F}$.

The rotationally averaged matrices $\RAp$ and $\RAm$ exhibit spherical symmetry, which means that they are diagonal in the multipolar basis of well defined total angular momentum, indexed by $j$, and angular momentum along the $\zhat$ axis, indexed by $m$. Moreover, for each $j$ subspace, the diagonal elements are equal for all $m\in [-j\ldots j]$ (see \cite[Eq. 7.5-13]{Tung1985}). The structure of the matrices is hence
\begin{equation}
	\label{eq:diagonal}
	\begin{bmatrix}
		c_1\Eye{3}&\Zero{3}{5}&\ldots&\ldots\\
		\Zero{5}{3}&c_2\Eye{5}&\ldots&\ldots\\
		\vdots&\vdots&\ddots&\vdots
	\end{bmatrix}.
\end{equation}
We now need the expansion of the circularly polarized incident plane waves $\vec{\mu}_\pm(\zhat)$ into regular multipolar fields. It can be found for example in \cite[Eq. (10.55)]{Jackson1998}. In the notation used in this paper it reads:
\begin{equation}
	\label{eq:jackjack}
	\mathbf{E}(\rr)^\omega_{\pm}=\sum_{j=1}^\infty i^j\sqrt{(4\pi)(2j+1)}\left[\MM_{jm=\pm 1}(\rr)\pm\NN_{jm=\pm 1}(\rr)\right].
\end{equation}
For a plane wave of helicity $\pm1$ the coefficients are zero except in the positions corresponding to angular momentum $m=\pm 1$. In order to bring Eq. (\ref{eq:jackjack}) to our conventions, we first note that, as per \cite[Eq. (10.46)]{Jackson1998}, it corresponds to the expansion of $\mathbf{E}(\rr)=(\xhat\pm i\yhat)\exp(ikz)$, whose polarization vector is a factor of $\sqrt{2}$ larger than its corresponding unitary vector. After we divide Eq. (\ref{eq:jackjack}) by $\sqrt{2}$, we can take this factor into the multipolar functions to obtain
\begin{equation}
	\frac{1}{\sqrt{2}}\left[\MM_{jm=\pm 1}(\rr)\pm\NN_{jm=\pm 1}(\rr)\right],
\end{equation}
which we manipulate to match the definitions of multipolar fields of well defined helicity $\lambda=\pm1$ in Eq. (\ref{eq:change})
\begin{equation}
	\frac{\lambda}{\sqrt{2}}\left[\NN_{jm=\lambda}(\rr)+\lambda\MM_{jm=\lambda}(\rr)\right]=\lambda \mathbf{A}^\omega_{jm=\lambda,\lambda}(\rr),
\end{equation}
It then follows that the expansion coefficients $\vec{\mu}_{\pm}^\omega(\vhat)$ that we are looking for are
\begin{equation}
	\label{eq:coefs}
	\mu_{jm\lambda}^\omega(\zhat)=\lambda \left(i^j\right)\sqrt{(4\pi)(2j+1)}\delta_{\lambda m},
\end{equation}
where $\delta_{\lambda m}$ is the Kronecker delta.

We can now evaluate the quadratic forms in Eq. (\ref{eq:CD}). For $\lambda=1$, and recalling the diagonal structure of $\RAp$ from Eq. (\ref{eq:diagonal}), we can write 
\begin{equation}
	\label{eq:tr}
	\begin{split}
		&\mupvz^\dagger\RAp\mupvz=\\
		&4\pi\Tr{{\RAp}_{1}}+4\pi\Tr{{\RAp}_{2}}+\ldots+\\
		&4\pi\Tr{{\RAp}_{j}}+\ldots=4\pi\Tr{\RAp},
	\end{split}
\end{equation}
where $\Tr{\matr{F}}_{j}$ is the trace of the submatrix of $\matr{F}$ which connects multipoles of order $j$ at the input to multipoles of the same order at the output. Equation (\ref{eq:tr}) is reached by considering that, from Eq. (\ref{eq:diagonal}), the diagonal elements of $\RAp$ are $c_j=\Tr{\matr{F}}_{j}/(2j+1)$, and that the combined action of $\mupvz^\dagger$ and $\mupvz$ is to select a single element on the diagonal for each multipolar order, and, according to Eq. (\ref{eq:coefs}), multiply it by $|\lambda \left(i^j\right)\sqrt{(4\pi)(2j+1)}|^2$.

The final step is the realization that it is not necessary to perform the rotational averages of the matrices because:
\begin{equation}
	\label{eq:traces}
	\Tr{{\RAp}_{j}}=\Tr{{\matr{A}_+^\omega}_{j}},
\end{equation}
and hence
\begin{equation}
	\label{eq:moretraces}
	\Tr{{\RAp}}=\Tr{{\matr{A}_+^\omega}}.
\end{equation}
Equations (\ref{eq:traces}) and (\ref{eq:moretraces}) follow from the facts that the trace is a rotationally invariant quantity, and that rotations do not mix the submatrices corresponding to different values of $j$.

After collecting these results, we obtain that Eq. (\ref{eq:CD}) can be written as
\begin{equation}
	\label{eq:CDexplicit}
	\boxed{\text{CD}^\omega=4\pi\Tr{\matr{A}^\omega_+-\matr{A}^\omega_-}.}
\end{equation}
The quantity in Eq. (\ref{eq:CDexplicit}) can be seen as a differential absorption probability. We now show how to express it in the familiar CD units [liter/mol/cm]. We can achieve this in two steps: 1) Convert absorption probability to absorption cross-section, and 2) use the conversion factor between absorption cross-section and units of [liter/mol/cm]: $10 N_A/\log(10)$ where $N_A=6.0221409\times10^{23}$ is Avogadro's number. Step 1) can be achieved comparing the expression for the absorption cross-section of a sphere under a circularly polarized plane-wave illumination in \cite[Eq. (10.61)]{Jackson1998}
\begin{equation}
	\label{eq:xsecs}
	\sigma_{abs}^\omega=\frac{\pi}{2{(k^\omega)}^2}\sum_j (2j+1)\left(2-|\alphaj-1|^2-|\betaj-1|^2\right),
\end{equation}
with our expression for the absorption probability
\begin{equation}
	\label{eq:pabs1}
	\text{p}_{abs}^\omega=\vec{\mu}_{\lambda}^\omega(\zhat)^\dagger\matr{A}_{\lambda}^\omega \vec{\mu}_{\lambda}^\omega(\zhat).
\end{equation}
We start by exploiting the structure of the T-matrix of a sphere. First, it does not couple multipoles with different $j$ or different $m$. Second, for each multipolar order $j$, it does not depend on $m$. And third, in the basis of multipoles of well defined helicity [Eq. (\ref{eq:change})], the 2$\times$2 submatrices relating incident and scattered multipoles of equal $m$ and $j$ read 
\begin{equation}
	\label{eq:submat}
-\frac{1}{2}\begin{bmatrix}\aj+\bj&\aj-\bj\\\aj-\bj&\aj+\bj\end{bmatrix},
\end{equation}
where $\aj$ and $\bj$ are the electric and magnetic Mie coefficients of the sphere, respectively.

Using the definitions of $\matr{A}^{\omega}_{\lambda}$ in Eq. (\ref{eq:abs}), and Eqs. (\ref{eq:coefs}) and (\ref{eq:submat}), one readily reaches from Eq. (\ref{eq:pabs1})
\begin{equation}
	\label{eq:pabs}
	\text{p}_{abs}^\omega=2\pi \sum_j (2j+1)\left(\mathbb{R}\left\{\aj+\bj\right\}-|\aj|^2-|\bj|^2\right),
\end{equation}
where $\mathbb{R}\left\{\cdot\right\}$ takes the real part. We now need to relate the Mie coefficients $(\aj,\bj)$ to the $(\alphaj,\betaj)$ coefficients in Eq. (\ref{eq:xsecs}), which can be done using the expansion of the field scattered by a sphere upon illumination with a circularly polarized incident plane wave \cite[Eq. (10.57)]{Jackson1998}. Particularized for incident helicity $\lambda=1$, it reads:
{\small
\begin{equation}
	\label{eq:jackjackjack1}
	\begin{split}
		\mathbf{E}_{sc}^\omega(\rr)&=\frac{1}{2}\sum_{j=1}^\infty i^j\sqrt{(4\pi)(2j+1)}\times\\
															 &\left[\alpha^{+}_j{\hatMM_{jm=1}}(\rr)+\beta_j^{+}{\hatNN_{jm=1}}(\rr)\right],
	\end{split}
\end{equation}
}
where $\hatMM(\rr)$ and $\hatNN(\rr)$ are outgoing multipoles. We now change the basis, the position of the leading $1/2$ factor, and divide by the previously discussed $\sqrt{2}$ factor:
{\small
\begin{equation}
	\label{eq:jackjackjack}
	\begin{split}
		\mathbf{E}_{sc}(\rr)&=\sum_{j=1}^\infty i^j\sqrt{(4\pi)(2j+1)}\times\\
																			   &\left[\frac{\beta_j^{+}+\alpha^{+}_j}{4}\hatAjmp+\frac{\beta_j^{+}-\alpha^{+}_j}{4}\hatAjmm\right].
	\end{split}
\end{equation}
}
We keep Eq. (\ref{eq:jackjackjack}) for future use and we now use the relationship between the coefficients of the incident plane wave $\vec{\mu}_+^\omega(\zhat)$, and the scattered field $\vec{\rho}^\omega(\zhat)$ [Eq. (\ref{eq:helicity})]
\begin{equation}
\begin{bmatrix}\vec{\rho}_{+}^\omega\\\vec{\rho}_{-}^\omega\end{bmatrix}=\begin{bmatrix}\Tpp&\Tpmi\\\Tmip&\Tmimi \end{bmatrix}\begin{bmatrix}\vec{\mu}_{+}^\omega\\\vec{0}\end{bmatrix},
\end{equation}
and Eqs. (\ref{eq:coefs}) and (\ref{eq:submat}) to write:
\begin{equation}
\label{eq:fff}
	\begin{split}
		\rho^\omega_{jm+}&=-\frac{(\aj+\bj)}{2}(i)^j\sqrt{(4\pi)(2j+1)}\delta_{1m},\\
	\rho^\omega_{jm-}&=-\frac{(\aj-\bj)}{2}(i)^j\sqrt{(4\pi)(2j+1)}\delta_{-1m}.\
	\end{split}
\end{equation}
Comparing Eqs. (\ref{eq:jackjackjack}) and (\ref{eq:fff}) shows that:
\begin{equation}
	\alphajp=-2\bj,\ \betajp=-2\aj,
\end{equation}
and allows to change Eq. (\ref{eq:xsecs}) into:
\begin{equation}
	\label{eq:sigma2}
	\sigma_{abs}^\omega=\frac{2\pi}{{(k^\omega)}^2}\sum_j (2j+1)\left(\mathbb{R}\left\{\aj+\bj\right\}-|\aj|^2-|\bj|^2\right).
\end{equation}
The same result is obtained using the opposite circular polarization.

Now, we note that the definition of the absorption cross-section includes the division by the incident flux. This means in particular that the $1/\sqrt{2}$ factor that we have been compensating for does not change it. We can hence, finally, compare Eqs. (\ref{eq:sigma2}) and (\ref{eq:pabs}) to reveal that
\begin{equation}
	\sigma_{abs}^\omega=\frac{\text{p}_{abs}^\omega}{{(k^\omega)}^2}.
\end{equation}
Therefore, in units of [liter/mol/cm], the frequency dependent circular dichroism as a function of the T-matrix reads:
\begin{equation}
	\overline{\text{CD}}^\omega=\frac{10N_A}{\log(10)}\frac{4\pi}{{(k^\omega)}^2}\Tr{\matr{A}^\omega_+-\matr{A}^\omega_-}.
\end{equation}

\section{Parity to helicity change\label{app:pth}}
The change of basis of Eq. (\ref{eq:change}) 
	{\small
\begin{equation}
		\Ajmp(\rr)=\frac{\NN_{jm}(\rr)+\MM(\rr)}{\sqrt{2}},\ \Ajmm(\rr)=\frac{\NN_{jm}(\rr)-\MM_{jm}(\rr)}{\sqrt{2}}, 
\end{equation}
}
induces the relationships
\begin{equation}
	\begin{split}
		\sqrt{2}\rho_{jm\pm}^\omega&=c_{jm}^\omega\pm d_{jm}^\omega\\
		\sqrt{2}\mu_{jm\pm}^\omega&=a_{jm}^\omega\pm b_{jm}^\omega\\
	\end{split}
\end{equation}
and
\begin{equation}
2\begin{bmatrix}\Tpp&\Tpmi\\\Tmip&\Tmimi \end{bmatrix}=\begin{bmatrix}\matr{I}&\matr{I}\\\matr{I}&-\matr{I} \end{bmatrix}\begin{bmatrix}\Tnn&\Tnm\\\Tmn&\Tmm\end{bmatrix}\begin{bmatrix}\matr{I}&\matr{I}\\\matr{I}&-\matr{I} \end{bmatrix},
\end{equation}
where $\matr{I}$ is the identity matrix.

\end{document}